\documentclass[a4paper]{article}
\usepackage{epsfig}
\newcounter{saveeqn}%
\newcommand{\alpheqn}{\setcounter{saveeqn}{\value{equation}}%
\stepcounter{saveeqn}\setcounter{equation}{0}%
\renewcommand{\theequation}
	{\mbox{\arabic{saveeqn}\alph{equation}}}}%
\newcommand{\reseteqn}{\setcounter{equation}{\value{saveeqn}}%
\renewcommand{\theequation}{\arabic{equation}}}%

\begin{document}

\author{Norbert Schorghofer\thanks{Email: norbert@phy.cuhk.edu.hk}\\
Department of Physics,
Chinese University of Hong Kong\\
Shatin, N.T., Hong Kong}
\title{Universality of probability distributions\\among two-dimensional turbulent flows}
\date{July 1, 1999}
\maketitle

\begin{abstract}
We study statistical properties of two-dimensional turbulent flows.  Three systems are considered: the Navier-Stokes equation, surface quasi-geostrophic flow, and a model equation for thermal convection in the Earth's mantle.  Direct numerical simulations are used to determine 1-point fluctuation properties.  Comparative study shows universality of probability density functions (PDFs) across different types of flow.  Especially for the derivatives of the ``advected'' quantity, the shapes of the PDFs are the same for the three flows, once normalized by the average size of fluctuations.  Theoretical models for the shape of PDFs are briefly discussed.
\end{abstract}

\vspace{0.5cm}

\twocolumn

The central idea of classical turbulence theory is that certain statistical properties in turbulent flow are independent of the details of the flow, like its boundaries, dissipation mechanism, and the kind of forcing, as long as the Reynolds number is sufficiently high.  In this sense turbulent flow would be universal.  Here we shall investigate independence not of boundary conditions, forcing etc., but look for universality across {\it equations}.  This is not a crazy idea at all and has long been demonstrated for other classes of partial differential equations.  Many partial differential equations with wave-like behavior, possessing common symmetries, exhibit identical fluctuation properties, once normalized by their standard deviations.  (The Schr\"odinger equation belongs to this class, as formulated in the well-known Bohigas-Giannoni-Schmit conjecture).  In the present paper it is demonstrated that three different hyperbolic partial differential equations (which all possess the same symmetries) exhibit, under analogous conditions, the same statistics for their fluctuations, once normalized by their standard deviations.

\alpheqn
The three flows are described by advection-diffusion equations
\begin{equation}
{\partial \theta \over \partial t}+ \vec v \cdot\nabla \theta = D \nabla^2 \theta +f.
\label{alpha1}
\end{equation}
The scalar quantity advected is $\theta$.  The forcing $f$ supplies the energy dissipated via a dissipation constant $D$.  The velocity $v$ is a function of $\theta$, best written in fourier space,
\begin{equation}
\hat {\vec v}(\vec k)=i {\vec k \times \hat\theta(\vec k) \over |\vec k|^\alpha}.
\label{alpha2}
\end{equation}
The two-dimensional cross product $\vec k\times \hat\theta$ is to be understood as a vector of length $|\vec k\hat\theta|$ and direction perpendicular to $\vec k$.  It follows that $\nabla \cdot \vec v = 0$.
\reseteqn

Different values of $\alpha$ correspond to different flows \cite{con94c}.  The two-dimensional Navier-Stokes equation is $\alpha=2$ and $\theta$ corresponds to the vorticity $\nabla \times \vec v$.
The surface quasi-geostrophic equation, $\alpha=1$, is a special case of the important quasi-geostrophic equation that describes flow of a shallow layer on a rotating sphere, as relevant for planetary atmospheres and oceans \cite{hel94}.  In this case, $\theta$ is physically interpreted as temperature, which drives the flow through its buoyancy effect.  The third equation considered is $\alpha=3$ which also appears in geophysical context as a limiting case of a shallow flow on a rotating sphere with uniform internal heating \cite{wei89}.  Also here, $\theta$ is a temperature.

Other values of $\alpha$, integer or not, could be considered, but this is not done here.  Finite-time singularities develop for $\alpha<1$, but not for $\alpha>1$.  For the border case of $\alpha=1$ no corresponding analytical proof has been achieved, but numerical evidence suggests that no finite-time singularities occur \cite{oy97,sch98a}.

Multiplying eqn. (\ref{alpha1}) with $\theta$ and averaging over space with periodic boundary conditions yields
\begin{equation}
{\partial\over \partial t}\left<\theta^2\right> = -D\left<|\nabla \theta|^2\right> +\left<f\theta\right>.
\end{equation}
Consequently the left hand side of eqn. (\ref{alpha1}) conserves $\left<\theta^2\right>$ for all $\alpha$.  For $\alpha=2$ also $\left<\vec v^2\right>$ is a conserved quantity, while  $\left<\vec v^2\right>\equiv\left<\theta^2\right>$ for $\alpha=1$, and  $\left<\vec v^2\right>$ is not conserved for $\alpha=3$.
Equation (1) is invariant under $r \to -r$ as well as the set of simultaneous transformations $r \to r\lambda$, $t \to t \lambda^2$, $\theta \to \theta/\lambda^\alpha$, and $f\to f/\lambda^{\alpha+2}$ (this is essentially the Reynolds number invariance).  The family of flows with different $\alpha$ has been named $\alpha$-turbulence \cite{hel94}, although this term appears in the literature also for other kinds of flow.

The flow is simulated in a doubly periodic square box.  Forcing acts on large scales, $4\le |k|<6$, with constant amplitude but random phases renewed at each time step.  (The time step is constant).  Two-dimensional turbulent flows produce vortices that merge and grow ever larger.  These vortices must be destroyed in order to reach an equilibrium state.  This is done by adding a large scale dissipation $-\gamma\theta$, with $\gamma$ small, to the right-hand side of eqn. (\ref{alpha1}), restricted to $0<|k|\le3$.  All these conditions are designed to produce isotropic and homogeneous flow.  The simulations are carried out with a pseudo-spectral method over long time periods using 4-th order Runge-Kutta integration.  A mild spectral filter has been used, without complete dealiasing, since it is not clear whether complete dealiasing improves or worsens the quality of simulations.

The aforementioned invariance naturally defines a Reynolds number for flow of any $\alpha$ as $Re=UL/D$, where $U$ and $L$ are a velocity and length scale respectively.  We choose $U=\sqrt{\left<\vec v^2\right>}$ and $L=1$ for a large-scale Reynolds number.  With this definition the maximum Reynolds numbers achieved are on the order of several thousands on a 1024x1024 grid for each of the three flows.

In this paper only 1-point probability density functions (PDFs) are studied.  First, the Navier-Stokes equation  ($\alpha=2$) is treated, which is important by itself and also exemplifies the variations and dependencies in the PDFs within one equation.  In the later part equations with different values of $\alpha$ are compared with each other, which is the central concern of this paper.


The PDFs are obtained from spatial snapshots of the flow field and then averaging over 8-24 such ensembles.   
Some of the PDFs shown are scaled by their average fluctuation, defined as
\begin{equation}
\sigma=\int{dx|x|P(x)}.
\end{equation}
The integral is over all $x$.  Instead of the first moment the standard deviation could have been used equally well.
All PDFs of the Navier-Stokes equation presented here agree with the ones reported from recent simulations by Takahashi and Gotoh \cite{tg96} at higher Reynolds numbers.

\begin{figure}[!ht]
a)\epsfxsize=8cm\epsfbox{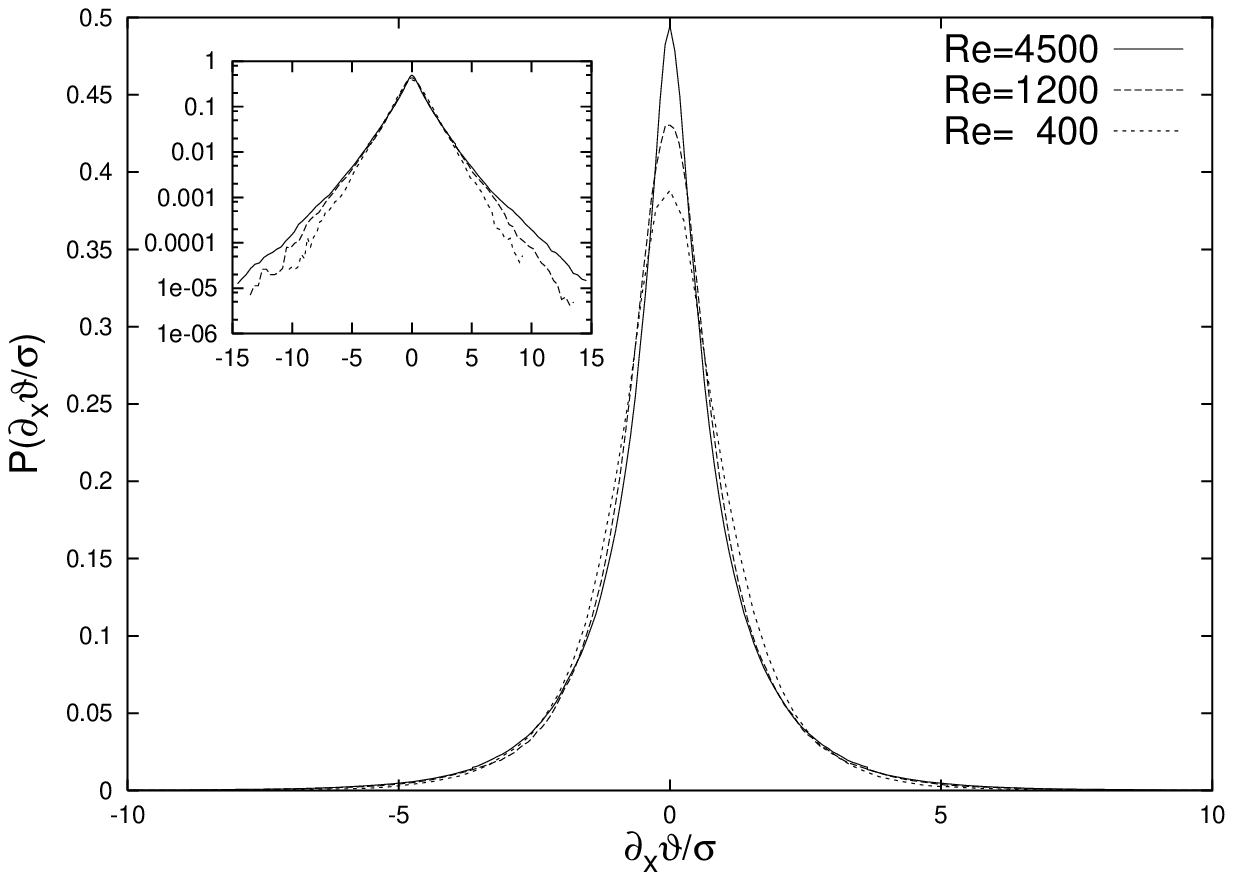}\\
b)\epsfxsize=8cm\epsfbox{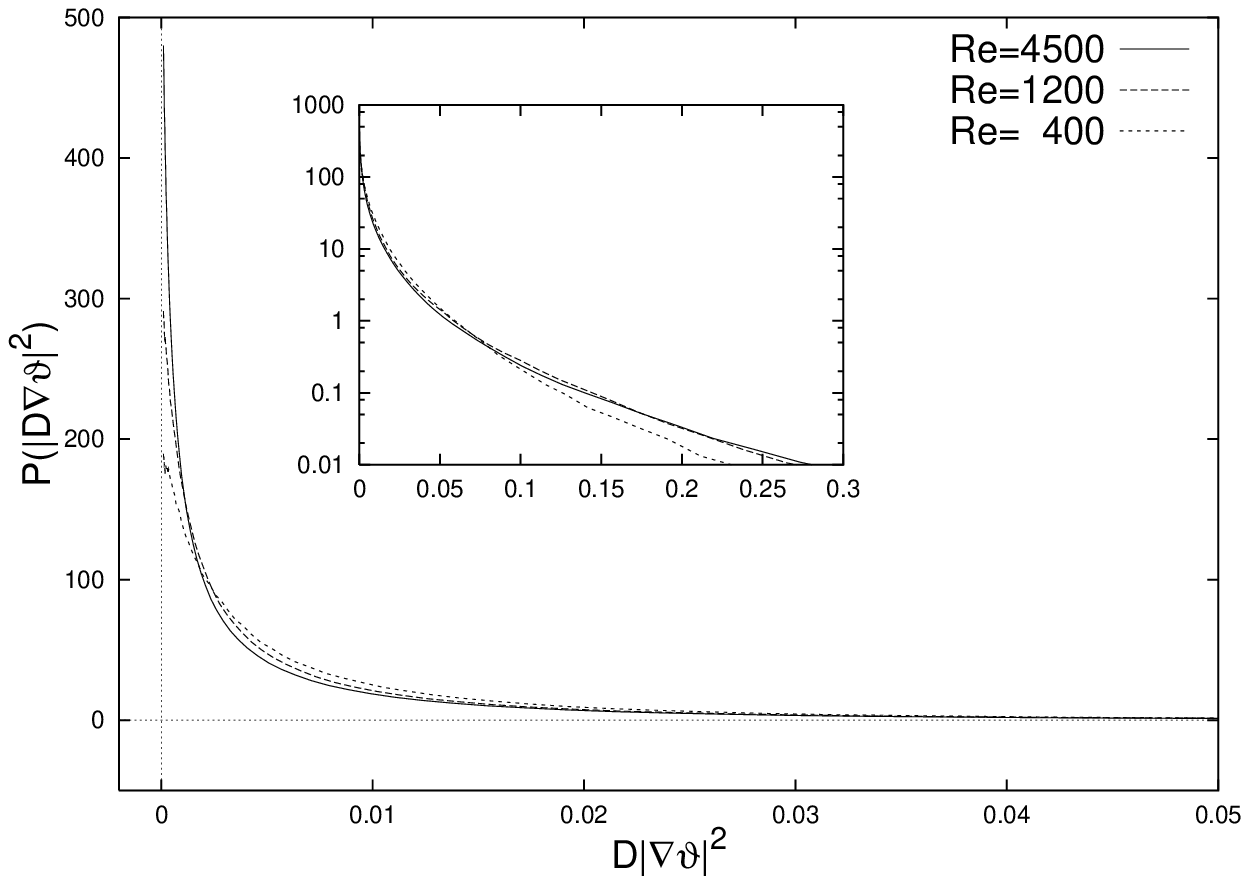}
\caption{Probability density functions for the Navier-Stokes equation at different Reynolds numbers  a) vorticity derivative $\partial_x\theta$ and b) vorticity dissipation $D|\nabla\theta|^2$.  Many small fluctuations account for most of the dissipation.  The insets show the same data on a logarithmic scale.
\label{reynolds}}
\end{figure}

Fig. \ref{reynolds} shows PDFs for different Reynolds numbers.  In Fig.  \ref{reynolds}a we see similar but not at all identical shapes for the PDFs, a behavior representative for the PDFs of other quantities as well.

\begin{figure}[!ht]
a)\epsfxsize=8cm\epsfbox{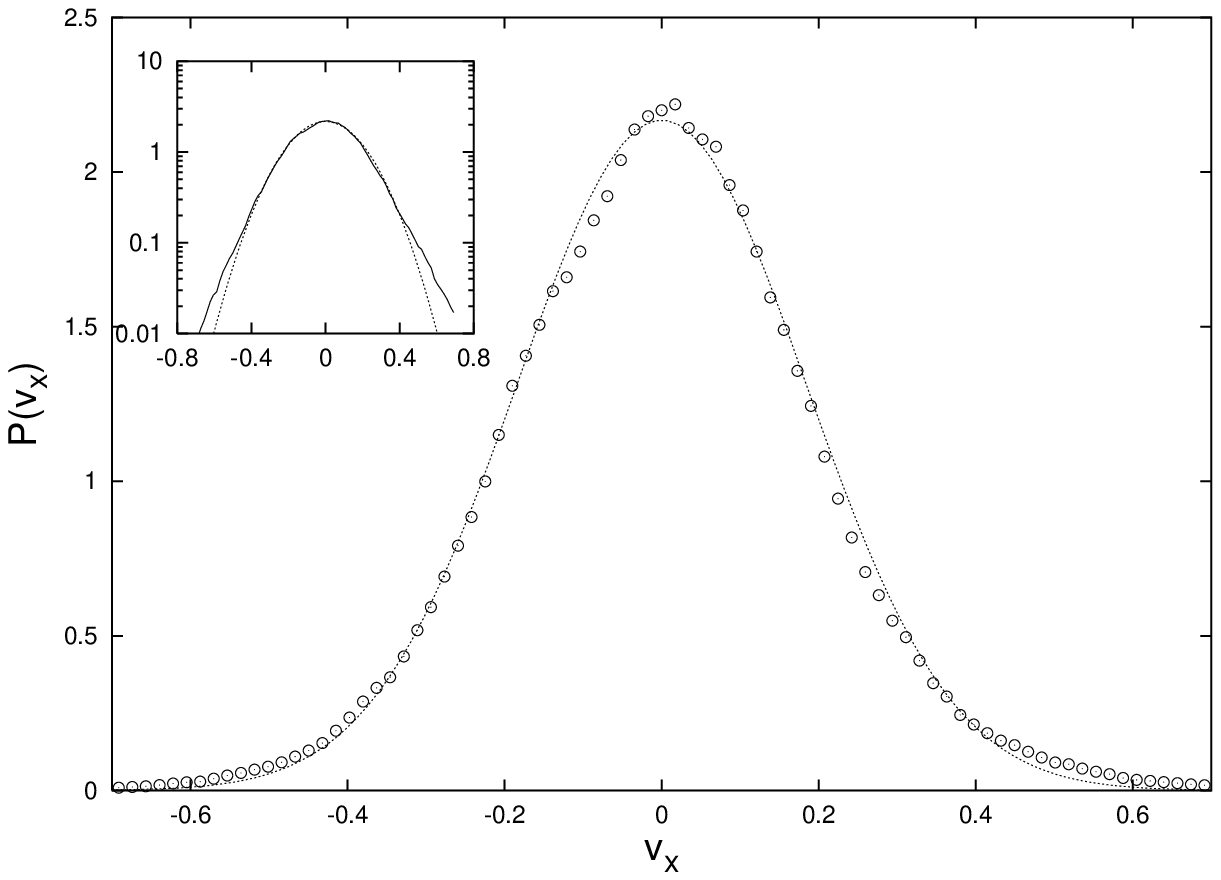}\\
b)\epsfxsize=8cm\epsfbox{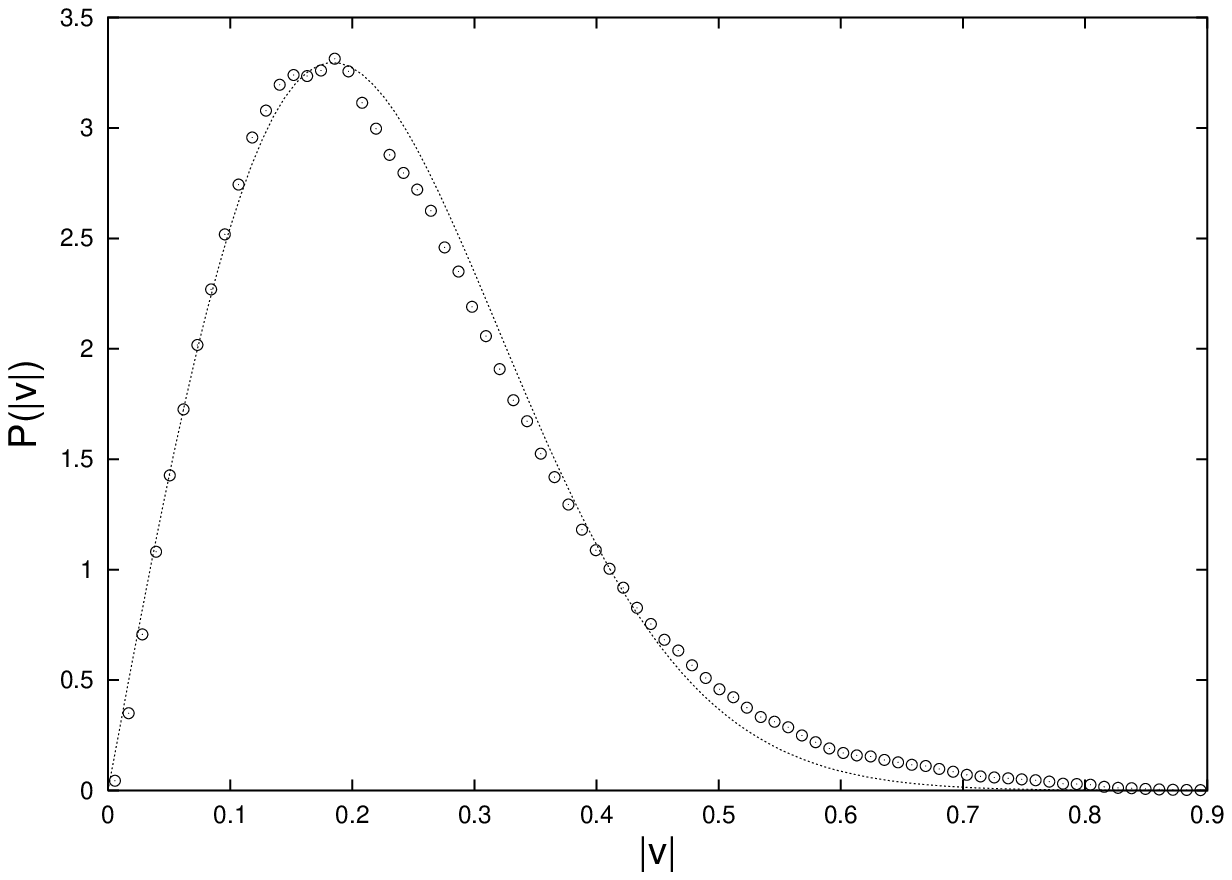}
\caption{Probability density functions for velocities of the Navier-Stokes at Re=4500.  a) velocity component $v_x$ and b) absolute value of velocity $|\vec v|$. The dotted lines are theoretical fits (Gaussian and Maxwellian).  
\label{velocities}}
\end{figure}

According to Fig. \ref{velocities}a velocity components are distributed Gaussian.  The PDFs for $v_x$ and $v_y$  are almost identical, as must be true for isotropic turbulence.  The tails show the same behavior as seen in decaying two-dimensonal turbulence.  For a detailed discussion of this issue see \cite{jim96}, where the tails are explained from the influence of large vortices, which cause the largest velocity gradients.
If the two velocity components are statistically independent of each other, then the PDF of the absolute value of $v$ should be a two-dimensional Maxwell distribution 
$$P(x)={x\over s^2}\exp{\left(-{x^2\over 2s^2}\right)}.$$  
The parameter $s$ is thereby the standard deviation of the Gaussian distribution for $v_x$.  The Maxwell distribution plotted as dotted curve in Fig. \ref{velocities}b, hence contains no free parameter.  It is a good first-order approximation.

As a matter of space not all PDFs can be presented here.  The scalar (vorticity) is Gaussian in the center.  The velocity derivatives are also Gaussian.  This is particularly striking, since velocity derivatives of decaying turbulence are not Gaussian \cite{jim96}.  Their core behaves much more like a Cauchy distribution
$$P(x)={1\over\pi}{c \over c^2+x^2},$$
which has an inflection point even on a logarithmic plot.  A Cauchy distribution follows theoretically from a ``dilute gas'' of point vortices of equal strength that move randomly, see \cite{jim96}.  For forced two-dimensional turbulence $\partial_x\theta$ (and $\nabla^2\theta$) could be interpreted as Cauchy distribution, but point vortices cannot make any sensible predictions on vorticity derivatives.  We shall return to velocity derivatives later.


PDFs in $\alpha$-turbulence have been previously reported in \cite{pie94b} (Fig. 7), where a ``remarkable similarity'' of the PDFs for $\theta$ has been pointed out.

\begin{figure}
a)\epsfxsize=8cm\epsfbox{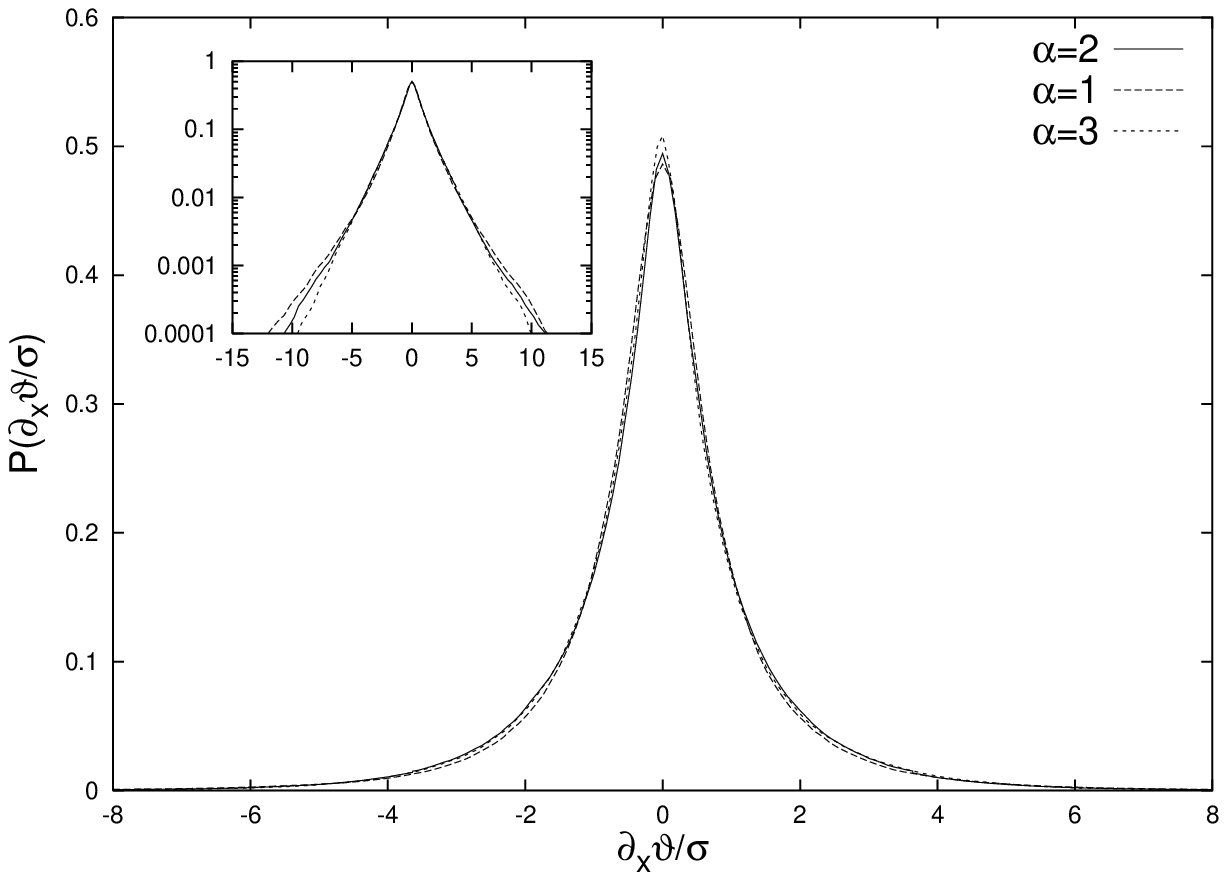}\\
b)\epsfxsize=8cm\epsfbox{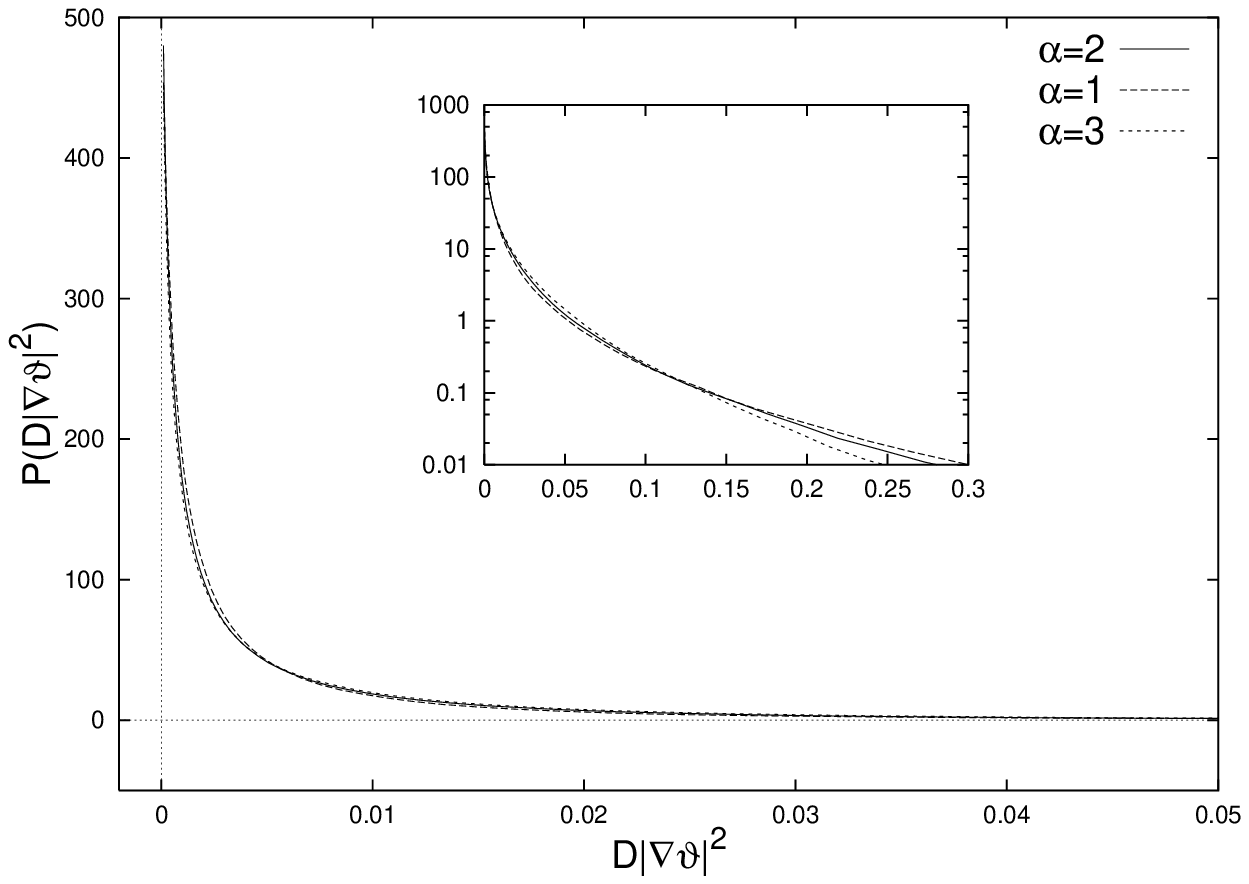}\\
c)\epsfxsize=8cm\epsfbox{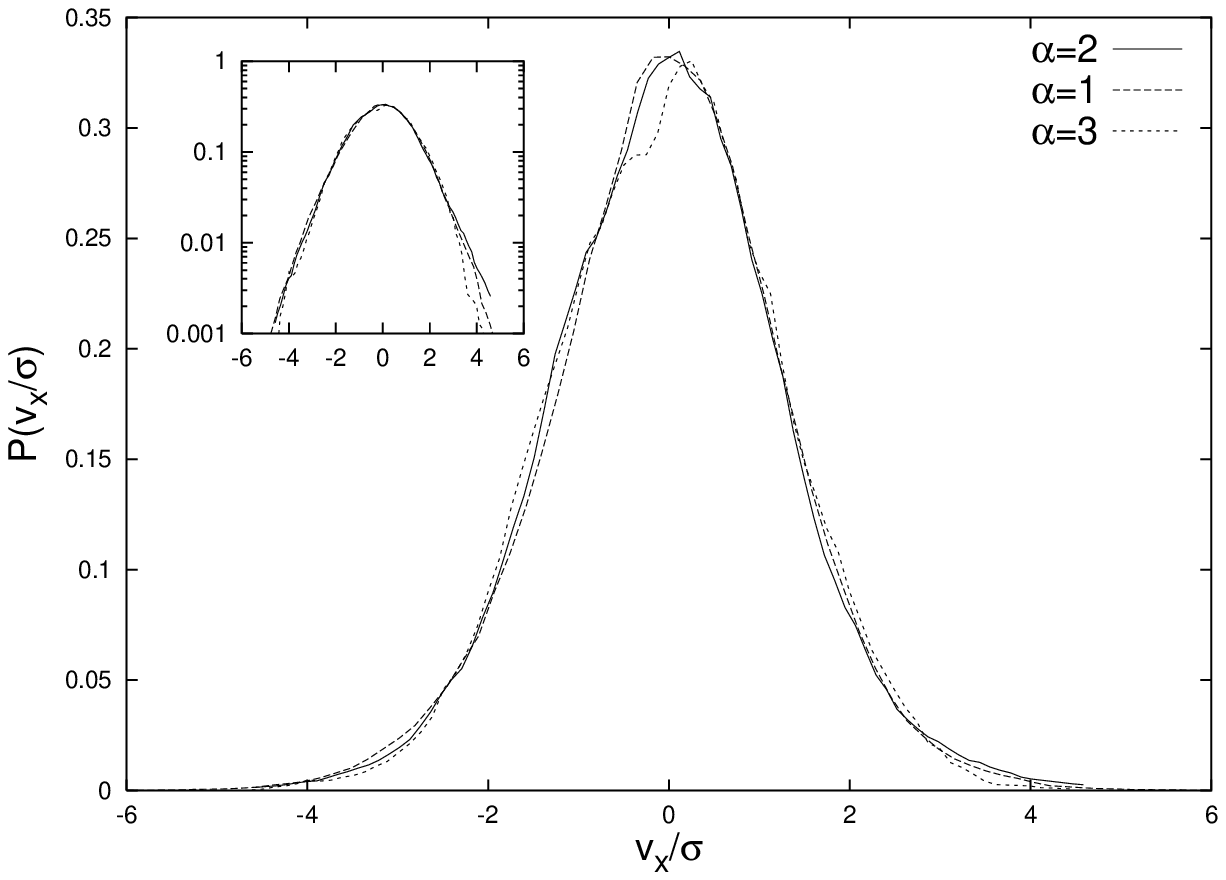}
\caption{The main result.  Probability density functions for different $\alpha$.  The flows have the same forcing and similar Reynolds number.  a) the gradient $\partial_x\theta$ b) the scalar dissipation $D|\nabla\theta|^2$, and c) the velocity component $v_x$.  The PDFs are divided by their average fluctuation $\sigma$, except for b, where this is not necessary.  The shape of the PDFs is independent of the type of flow.
\label{universality}} 
\end{figure}

Fig. \ref{universality} shows PDFs for different types of flow.  In each figure the PDFs of all three flows are shown simultaneously, and the differernt figures show scalar derivatives $\partial_x\theta$, scalar dissipation $D|\nabla\theta|^2$, and velocity components $v_x$.  Apparently the PDFs for the different flows are the same.  The agreement is for small as well as large fluctuations up to several standard deviations.  The very largest fluctuations are inevitably undersampled, which accounts for deviations in the tails, which are likely to fall within measurement errors.

PDFs by themselves are known not to be particularly robust or universal, and it is hence important to compare flows under analogous conditions.  The Reynolds numbers in the simulations for $\alpha=1,2,3$ were $Re=$ 3900, 4500, 4200 respectively.  The differences do not appear significant.  Comparisons at a lower Reynolds number (around 1300) yield the same universalities.

Also $\nabla^2\theta$ (not shown) overlaps with the same accuracy as $\partial_x\theta$ and $|\nabla\theta|^2$.  But not all PDFs overlap as accurately as the derivatives of $\theta$.  The deviations in Fig. \ref{universality}c are somewhat larger.  Other PDFs show even larger deviations, but in none of the PDFs investigated was there any drastic difference.  These deviations could lie within measurement errors.  It turns out that quantities with larger deviations also show greater variability within the same equation for different Reynolds number.  It appears that some PDFs simply require longer averaging than others.  It is consistent with the data that all 1-point PDFs are identical, but it is also consistent that some are not.  At the very least, PDFs for $\partial_x\theta$, $|\nabla\theta|^2$, $\nabla^2\theta$, $v_x$ (and $\partial_y\theta$, $v_y$) agree within several standard deviations.

We note (from Figures \ref{reynolds}a and \ref{universality}a) that the variation in the PDFs for different $\alpha$ is smaller than the variation with Reynolds number.  It is astonishing that universality across equations is more robust than within the same equation at different Reynolds numbers.

The three quations describe physically quite different flows and also their mathemtical properties are diverse.  For example, the velocity is a local function of the scalar only for one of the equations, and the kinetic energy is conserved in only two of the three flows.  Inspite of these vast differences, their fluctuation properties differ only in absolute size.

PDFs of velocities on the ocean surface have recently been measured \cite{lsg98}.  Although none of the equations described here is directly applicable to this situation, the velocity components and their derivatives (the latter are not shown here) apparently show the same behavior as forced $\alpha$-turbulence.  Universality across equations might explain the observed PDFs.

Several, if not all, probability density functions of three different flows, described by hyperbolic partial differential equations with reflection symmetry and Reynolds number invariance, have identical shapes.  This clearly demonstrates that fluctuation properties can be explained in terms of statistical considerations only, and do not (or only very marginally) involve the dynamics of the flow.

\vspace{0.5cm}\noindent
{\bf Acknowledgments:} It is a pleasure to thank Alexander Bershadskii, Jeff Chasnov, Emily Ching, and Toshiyuki Gotoh for very helpful discussions.  Frank Ng and George Yuen from the High Performance Computing System at the Chinese University of Hong Kong provided indespensable technical help.  This work was supported by a postdoctoral fellowship from the Chinese University of Hong Kong and a grant from the Research Grants Council of the Hong Kong Special Administrative Region, China (RGC Ref. No. CUHK4119/98P).

\bibliographystyle{unsrt}

\end{document}